\documentclass[prb,aps,onecolumn,showpacs,floats,a4paper,times,12pt]{revtex4}
\usepackage{epsfig,bm,epsf,graphics}
\begin{document}
\draft
\title{SPIN TRANSPORT IN MAGNETICALLY ORDERED SYSTEMS: EFFECT OF THE LATTICE RELAXATION TIME}
\author{Y. Magnin, Danh-Tai Hoang,  and H. T. Diep\footnote{ Corresponding author, E-mail: diep@u-cergy.fr }}
\address{
Laboratoire de Physique Th\'eorique et Mod\'elisation,
Universit\'e de Cergy-Pontoise, CNRS, UMR 8089\\
2, Avenue Adolphe Chauvin, 95302 Cergy-Pontoise Cedex, France\\}

\begin{abstract}
Spin resistivity $R$ has been shown to result mainly from the scattering of itinerant spins with magnetic impurities and lattice spins.   $R$ is proportional to the spin-spin correlation so that its behavior is very complicated near and at the magnetic phase transition of the lattice spins.   For the time being there are many new experimental data on the spin resistivity going from  semiconductors to superconductors.  Depending on materials, various behaviors have been observed.  There is however no theory so far which gives a unified mechanism for spin resistivity in magnetic materials. Recently,  we have showed Monte Carlo results for different systems.  We found that the spin resistivity is very different from one material to another.  In this paper, we show for the first time how the dynamic relaxation time of the lattice spins affects the resistivity of itinerant spins observed in Monte Carlo simulation.
\end{abstract}

\maketitle

\section{Introduction}

The resistivity is an important subject in condensed-matter physics. It has been studied experimentally and theoretically already with old classical physics.  However, only from the fifties, with notions borrowed from microscopic modern physics that the resistivity has been viewed as a consequence of microscopic mechanisms which govern physical behaviors of materials through which conduction electrons travel.  In this paper, we are interested in the resistivity caused by magnetic scattering of itinerant electronic spins by localized lattice spins in magnetic materials (ferromagnets and anfiferromagnets). The resulting resistivity is called hereafter "spin resistivity" which is to be distinguished from the resistivity due to spin-independent scattering, for example by phonons and non magnetic impurities.

The spin resistivity, $R$,  was shown to depend on the spin-spin correlation in ferromagnetic crystals by de Gennes and Friedel\cite{DeGennes}, Fisher and Langer\cite{Fisher} among others.  At low temperatures ($T$), the spin-waves are shown to be responsible for the $T^2$ behavior of the spin resistivity in ferromagnets\cite{Kasuya,Turov}.  Note however that in these calculations the itinerant electrons have been considered as  free electrons interacting with the lattice spins, but there is no interaction between them. We have showed\cite{Magnin,Magnin2} that when an interaction between itinerant electrons is introduced,  the itinerant electrons can be crystallized at low $T$ giving rise to an increase of $R$ as $T\rightarrow 0$. Experimental data in various materials show this behavior\cite{Chandra,Du,McGuire,Lu,Santos}, but we would warn that there may be other mechanisms involved as well.
At the magnetic phase transition temperature $T_C$, the spin-spin correlation diverges in magnetic materials with a second-order phase transition. The theory of de Gennes-Friedel predicts that $R$ should show a divergent peak. However, experiments in various magnetic materials ranging from semiconductors to superconductors\cite{Chandra,Du,McGuire,Lu,Santos,Stishov,Stishov2,Xia,Wang-Chen,Li,Zhang} show indeed an anomaly at the transition temperature $T_C$, but the peak  is more or less rounded, not as sharp as expected from the divergence of the correlation length.  It has been shown in fact that\cite{Fisher,Kataoka} the form of the peak depends on the length of the correlation included in the calculation of $R$: if only short-range correlations are taken into account, then the peak is very rounded. A justification for the use of only short-range correlations comes from the fact that the mean free path of itinerant spins is finite at $T_C$.   When scattering is due to impurities, the peak has been shown to depend on the localization length\cite{Zarand}.  In the case of antiferromagnets, Haas has shown the absence of a resistivity peak\cite{Haas}.
Our recent works using Monte Carlo (MC) simulations have shown that there is indeed an anomaly at $T_C$ in various magnetic models from ferromagnets\cite{Magnin,Akabli,Akabli2}, antiferromagnets\cite{Magnin,Akabli3} to frustrated spin systems\cite{Magnin2,Hoang}.   The shape of the anomaly depends on many ingredients such as crystal structures, spin models, and interaction parameters.

In this paper, we will show  new results obtained by MC simulation when we take into account the temperature dependence of the relaxation time of localized lattice spins in the simulation.  We will show that this temperature dependence affects the shape of the peak in the phase transition region.

Section \ref{Model} is devoted to a description of the general model and the MC method.  We introduce in this section the temperature dependence of the relaxation time.   Results are shown and discussed in section  \ref{Res} for both ferro- and antiferromagnets in terms of critical slowing-down. Concluding remarks are given in section \ref{Concl}.

\section{Model and Method}\label{Model}

The model we use in our MC simulation is very general.  The itinerant spins move in a crystal whose lattice sites are occupied by localized spins.  The itinerant spins are assumed to be of Ising type, but the method of simulation can be used for other spin models\cite{Akabli3}. The localized spins may be of Ising, XY or Heisenberg models.  Their interaction is usually limited to nearest neighbors (NN) but this assumption is not necessary. It can be ferromagnetic or antiferromagnetic.

\subsection{Interactions}

We consider a thin film of  a given lattice structure where each lattice site is occupied by a spin.   The interaction between the lattice spins is limited to NN with
the following Hamiltonian :
\begin{eqnarray}
\mathcal{H}_l & = & -\sum_{(i,j)}J_{i,j}\vec{S}_{i}.\vec{S}_{j}\label{HamilR}
\end{eqnarray}
where $\vec S_i$ is an Ising spin whose values are $\pm 1$, $J_{i,j}$ the exchange integral between the NN spin pair $\vec S_i$ and $\vec S_j$.  Hereafter we take $J_{i,j}=J$ for all NN spin pairs, for simplicity. As a convention, ferromagnetic (antiferromagnetic) interaction has positive (negative) sign.   The system size is $N_x \times N_y \times N_z$ where $N_i(i=x,y,z)$ is the number of lattice cells in the $i$ direction. Periodic boundary conditions (PBC) are used in the $x$ and $y$ directions while the surfaces perpendicular to the $z$ axis are free. The film thickness is $N_z$.

We define the interaction between the itinerant spins and the localized lattice spins as follows
\begin{eqnarray}
\mathcal{H}_r & = & -\sum_{i,j}I_{i,j}\vec{\sigma}_i .\vec{S}_j\label{I}
\end{eqnarray}
where $\sigma_i$ is the Ising spin of the $i-th$ itinerant electron and $I_{i,j}$ denotes the interaction that
depends on the distance between an electron $i$ and the spin $\vec{S}_j$ at the lattice site $j$. For simplicity, we use the following interaction expression
\begin{eqnarray}
I_{i,j} & = & I_{0}e^{-\alpha r_{ij}}
\end{eqnarray}
where $r_{ij}=|\vec{r}_i-\vec{r}_j|$,  $I_0$ and $\alpha$ are  constants. In the same way, interaction between itinerant electrons is
defined by
\begin{eqnarray}
\mathcal{H}_m & = & -\sum_{i,j}K_{i,j}\vec{\sigma}_i .\vec{\sigma}_j\\
K_{i,j} & = & K_{0}e^{-\beta r_{ij}}\label{K}
\end{eqnarray}
with $K_{i,j}$ being the interaction that depends on the distance between electrons $i$ and $j$.
The choice of the constants $K_0$ and $\beta$ will be discussed below.

Dynamics of itinerant electrons is ensured by an electric field applied along the $x$ axis. Electrons  travel in the $x$ direction, leave the system at the end. The PBC on the $xy$ planes ensure that the electrons who leave the system at one end are to be reinserted at the other end. For the $z$ direction, we use the mirror reflection at the two surfaces.  These boundary conditions  are used in order to conserve the average density of itinerant electrons.  One has
\begin{eqnarray}
\mathcal{H}_E & = & -e\vec{\epsilon}.\vec{\ell}
\end{eqnarray}
where $e$ is the charge of electron, $\vec \epsilon $ the applied electric field and $\vec \ell$ the
displacement vector of an electron.

Since the interaction between itinerant electron spins is attractive,  we need to add a chemical
potential in order to avoid a possible agglomeration of electrons into some points in the crystal and to ensure a homogeneous spatial distribution of electrons during the simulation. The chemical potential term is given by
\begin{eqnarray}
\mathcal{H}_c & = & D\vec{\nabla}_rn(\vec{r})\label{pot}
\end{eqnarray}
where $n(\vec r)$ is the concentration of itinerant spins in the sphere of $D_2$ radius, centered at $\vec r$. $D$ is a constant parameter appropriately chosen.\\

\subsection{Simulation Method}

The procedure of our simulation can be split into two steps. The first step consists in equilibrating
the lattice at a given temperature $T$ without itinerant electrons.  When equilibrium is reached,
in the second step, we randomly add $N_0$ polarized itinerant spins into the lattice. Each itinerant
electron interacts with lattice spins in a sphere of radius $D_1$ centered at its position, and with other itinerant electrons in a sphere of radius $D_2$.

The procedure of spin dynamics is described as follows. After injecting $N_0$ itinerant
electrons in the equilibrated lattice, we equilibrate the itinerant spins using the following updating. We calculate the energy $E_{old}$ of an itinerant electron taking into account all interactions described above. Then we perform a trial move of length $\ell$ taken in an arbitrary direction with random modulus in the interval $[R_1,R_2]$ where $R_1=0$ and $R_2=a$ (NN distance),  $a$ being the lattice constant.  Note that the move is rejected if the electron falls in a sphere of radius $r_0$ centered at a lattice spin or at another itinerant electron. This excluded space emulates the  Pauli exclusion. We calculate the new energy $E_{new}$ and use the Metropolis algorithm to accept or reject the electron displacement.
We choose another itinerant electron and begin again this procedure.  When all itinerant electrons
are considered, we say that we have made a MC sweeping, or one MC step/spin. We have to repeat a large number of MC steps/spin to reach a stationary transport regime. We then perform the averaging to determine physical properties such as magnetic resistivity, electron velocity, energy etc. as functions of temperature.

We emphasize here that in order to have sufficient statistical averages on microscopic states of
both the lattice spins and the itinerant spins, we use the following procedure: after averaging  the resistivity over $N_1$ steps for "each" lattice spin configuration, we thermalize again the lattice with $N_2$ steps in order to take another disconnected lattice configuration. Then we take back the averaging of the resistivity for $N_1$ steps for the new lattice configuration. . We repeat this cycle  for $N_3$ times, usually several hundreds of thousands times.  The total MC steps for averaging is  about  $4\times 10^5$ steps per spin in our simulations.   This procedure reduces  strongly thermal fluctuations observed in our previous work\cite{Akabli2}.

Of course, the larger $N_1$ and $N_3$ are the better the statistics becomes.  The question is what is the correct value of $N_1$ for averaging with each lattice spin configuration at a given $T$?  This question is important because this is related to the relaxation time $\tau_L$of the lattice spins compared to that of the itinerant spins, $\tau_I$.  The two extreme cases are i) $\tau_L \simeq \tau_I$, one should take $
N_1=1$, namely the lattice spin configuration should change with each move of itinerant spins ii) $\tau_L \gg \tau_I$, in this case,  itinerant spins can travel in the same lattice configuration for many times during the averaging.

In order to choose a right value of $N_1$, we consider the following temperature dependence of $\tau_L$ in non frustrated spin systems.  The relaxation time is expressed in this case as\cite{Hohenberg}
\begin{equation}\label{tau}
\tau_L=\frac{A}{|1-T/T_C|^{z\nu}}
\end{equation}
where $A$ is a constant, $\nu$  the correlation critical exponent, and $z$ the dynamic exponent.  From this expression, we see that as $T$ tends to $T_C$, $\tau_L$ diverges.  In the critical region around $T_C$ the system encounters thus the so-called "critical slowing down": the spin relaxation is extremely long due to the divergence of the spin-spin correlation.  In our previous papers\cite{Akabli,Akabli2,Magnin,Magnin2,Hoang}, we did not take into account the temperature dependence of $\tau_L$.   We propose to study here the spin resistivity using   Eq. (\ref{tau}).

We define spin resistivity $\rho$ as :
\begin{eqnarray}
\rho & = & \frac{1}{n_{e}}
\end{eqnarray}
where $n_{e}$ is the number of itinerant electron spins crossing a  unit slice perpendicular to the $x$ direction per unit of time.

\subsection{Choice of parameters and units}\label{choice}
The spin resistivity is dominated by the two interactions Eqs. (\ref{I}) and (\ref{K}).  As said earlier, our model is very general. Several kinds of materials such as metals, semiconductors, insulating magnetic materials etc. can be studied with our model, provided an appropriate choice of the parameters. For example, non magnetic metals correspond to $I_{i,j}= K_{i,j}=0$ (free conduction electrons). The case of magnetic semiconductors corresponds to the choice of parameters $K_0$ and $I_0$ so as the energy of an itinerant electron due to the interaction $\mathcal{H}_r$ should be much lower than that due to $ \mathcal{H}_m$, namely itinerant electrons are more tightly bound to localized atoms.  Note that $\mathcal{H}_m$ depends on the concentration of itinerant spins: for example the dilute case yields a small $\mathcal{H}_m$.
We will show below results obtained for typical values of parameters which correspond more or less to semiconductors.  The choice of the parameters has been made after numerous test runs.  We describe the principal requirements which guide the choice:
i) We choose the interaction between lattice spins as unity, i. e. $|J|=1$,
ii) We choose interaction between an itinerant and its surrounding lattice spins so as its energy $E_i$ in the low $T$ region is the same order of magnitude with that between lattice spins. To simplify, we take $\alpha=1$.  This case corresponds more or less to a semiconductor, as said earlier,
iii) Interaction between itinerant spins is chosen so that this contribution to the itinerant spin energy is smaller than
$E_i$ in order to highlight the effect of the lattice ordering on the spin current. To simplify, we take $\beta=1$,
iv) The choice of $D$ is made in such a way to avoid the formation of  clusters of itinerant spins (agglomeration) due to their attractive interaction [Eq. (\ref{K})],
v) The electric field is chosen not so strong in order to avoid its dominant effect that would mask the effects of thermal fluctuations and of the magnetic ordering,
vi) The density of the itinerant spins is chosen in a way that the contribution of interactions between themselves is much weaker than $E_i$, as said above in the case of semiconductors.

A variation of each parameter respecting the above requirements does not change qualitatively the results shown below. Only the variation of $D_1$ in some antiferromagnets does change the results (see Ref. \cite{Magnin2}).

The energy is  measured in the unit of $|J|$. The temperature is expressed in the unit of $|J|/k_B$.  The distance ($D_1$ and $D_2$) is in the unit of $a$.

\section{Results and Discussion}\label{Res}
In this section, we show for comparison the results obtained with and without temperature dependence of the lattice relaxation time for both ferromagnets and antiferromagnets. In each case we use the same set of interaction parameters in order to outline the effect of the temperature-dependent relaxation time.

In this paper we use the lattice size $N_x=N_y=20$ and $N_z=8$ and we consider the body-centered cubic (BCC) lattice for illustration.  The lattice constant is $a$.
The spin resistivity is calculated with $N_0=(N_x\times N_y\times N_z)/2$ itinerant spins (one electron per  two lattice cells).  Except otherwise stated, we choose interactions $I_0=2$, $K_0=0.5$, $D_1=a$, $D_2=a$, $D=0.5$, $\epsilon=1$, $N_0=1600$, and $r_0=0.05a$.  A discussion on the effect of a variation of each of these parameters is given above.

Note that, due to the form of the interaction given by Eq. (\ref{K}),  the itinerant spins
have a tendency to form compact clusters to gain energy.  This tendency is neutralized by the concentration gradient term, i. e. a chemical potential, given by  Eq. (\ref{pot}).  The value of $D$ has to be chosen so as to avoid an agglomeration of itinerant spins.   This choice depends of course on the values of $D_1$ and $D_2$.  Examples have been shown elsewhere.\cite{Magnin2,Hoang}
For the temperature dependence of the lattice relaxation time $\tau_L$, we take $\nu=0.638$ (3D Ising universality) and $z=2.02$.\cite{Prudnikov}  By choosing $A=1$, we fix $\tau_L=1$ at $T=2T_C$ deep inside the paramagnetic phase far above $T_C$. This value is what we expect for thermal fluctuations in the disordered phase.

\begin{figure}[th]
\centerline{\psfig{file=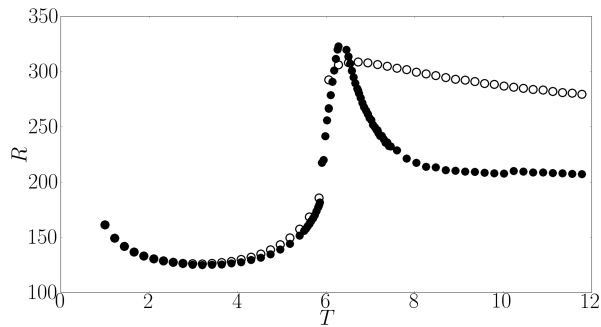,angle=0,width=9cm}}
\vspace*{8pt}
\caption{ BCC ferromagnetic thin film: Resistivity $R$ with temperature-independent relaxation (white circles) and temperature-dependent relaxation (black circles) in arbitrary unit versus temperature
$T$, in zero magnetic field, with electric field $\epsilon=1$, $I_0=2$, $K_0=0.5$.}\label{RX-F}
\end{figure}

Figure \ref{RX-F} shows the spin resistivity $R$ in a BCC ferromagnet.  Note that the transition temperature for this thin film of size $20\times20\times 8$ with Ising spins interacting via the NN coupling is $T_C\simeq 6.35$. Several remarks are in order:

i) The results obtained with and without temperature-dependent relaxation time  for $T<T_C$ coincide with each other

ii) At $T_C$, for the set of parameters used here, the results using the temperature-independent relaxation shows a broad maximum above $T_C$ while those using the temperature-dependent relaxation strongly decreases at $T_C$ giving rise to a sharp peak.

\begin{figure}[th]
\centerline{\psfig{file=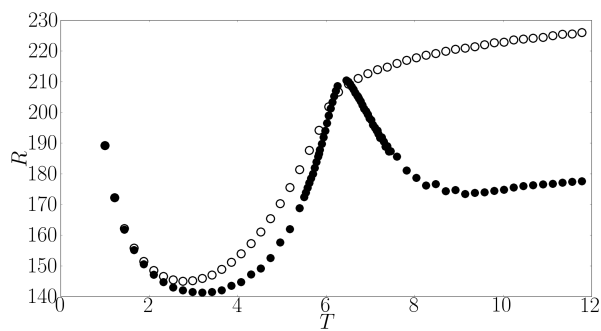,angle=0,width=9cm}}
\vspace*{8pt}
\caption{ BCC antiferromagnetic thin film: Resistivity $R$ with temperature-independent relaxation (white circles) and temperature-dependent relaxation (black circles) in arbitrary unit versus temperature
$T$, in zero magnetic field, with electric field $\epsilon=1$, $I_0=2$, $K_0=0.5$.}\label{RX-AF}
\end{figure}

We show now in Fig. \ref{RX-AF}  the spin resistivity $R$ in a BCC antiferromagnet.  Note that the transition temperature is the same as that of the ferromagnet counterpart shown above.  Here we observe that $R$ in the case of temperature-dependent relaxation is lower than that in the case of temperature-independent one in the whole temperature range.  Note that the value of the peak is much smaller here than in the ferromagnet case.

%\begin{figure}[th]
%\centerline{\psfig{file=fig2bis.eps,angle=-90,width=9cm}}
%\vspace*{8pt}
%\caption{Resistivity $R$ of the FCC ferromagnet in arbitrary unit versus temperature
%$T$ for several values of $I_0$: 2 (black circles), 1 (void circles), 0.5 (black triangles).  Other parameters: $N_x=N_y=20$, $N_z$=8, $\epsilon=1$, $K_0=0.5$, $D=0.5$, $D_1=1$. \label{R}}
%\end{figure}

%\begin{figure}[th]
%\centerline{\psfig{file=fig5bis.eps,angle=-90,width=9cm}}
%\vspace*{8pt}
%\caption{ SC AF case. Resistivity $R$ in arbitrary unit versus temperature
%$T$, in zero magnetic field, with electric field $\epsilon=1$, $I_0=K_0=0.5$.}\label{R-AF}
%\end{figure}

\begin{figure}[th]
\centerline{\psfig{file=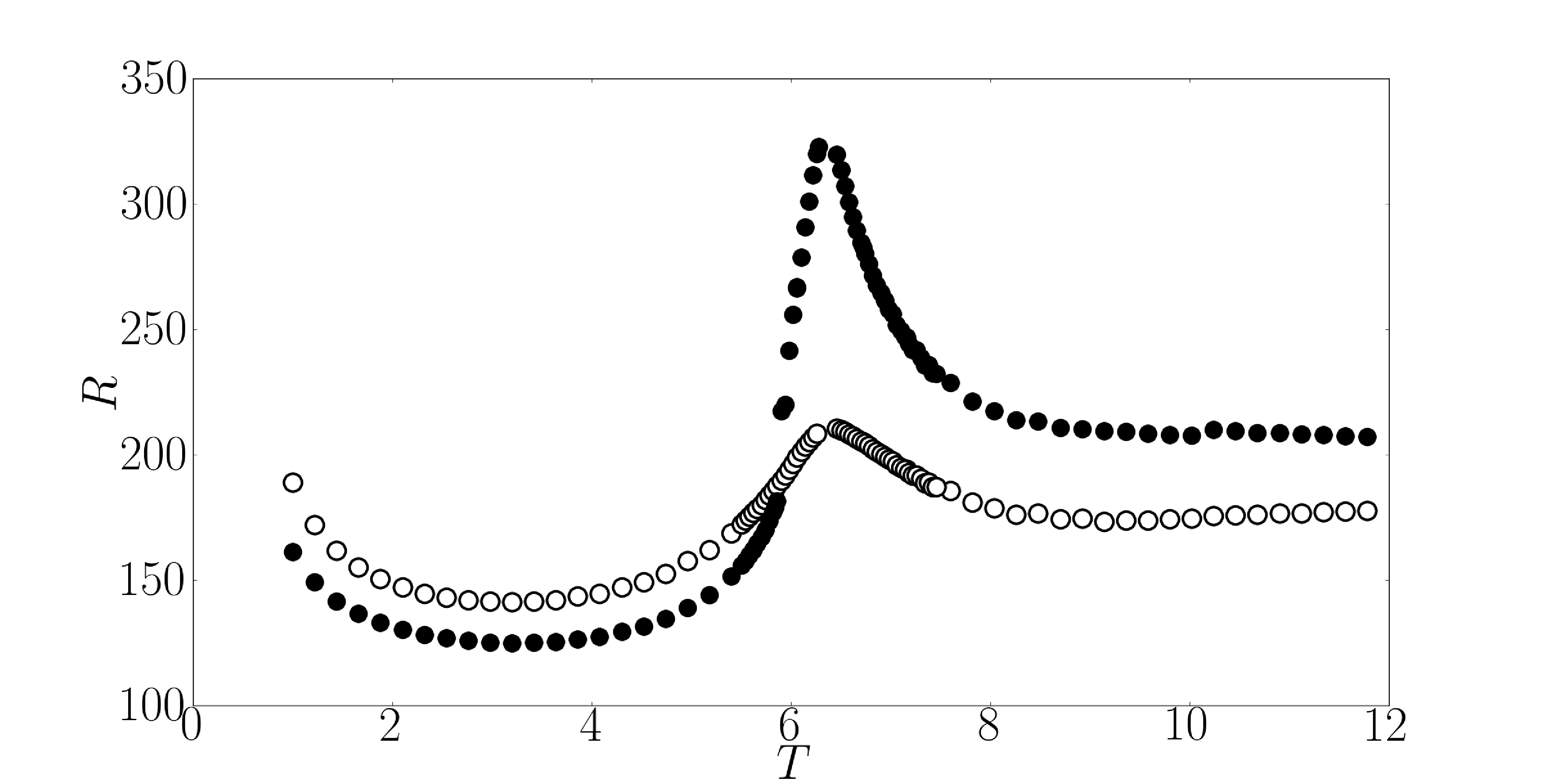,angle=0,width=9cm}}
\vspace*{8pt}
\caption{ BCC ferromagnetic and antiferromagnetic films:  Resistivity $R$ with temperature-dependent relaxation for ferro- (black circles) and antiferromagnet (white circles) in arbitrary unit versus temperature
$T$, in zero magnetic field, with electric field $\epsilon=1$, $I_0=2$, $K_0=0.5$.}\label{RX-F-AF}
\end{figure}

We show in Fig. \ref{RX-F-AF} the two curves of  ferromagnet and antiferromagnet with $T$-dependent relaxation time.  We observe here that below $T_C$, the resistivity of antiferromagnet is higher than that of ferromagnet, while for $T>T_C$ the reverse is true.

\begin{figure}[th]
\centerline{\psfig{file=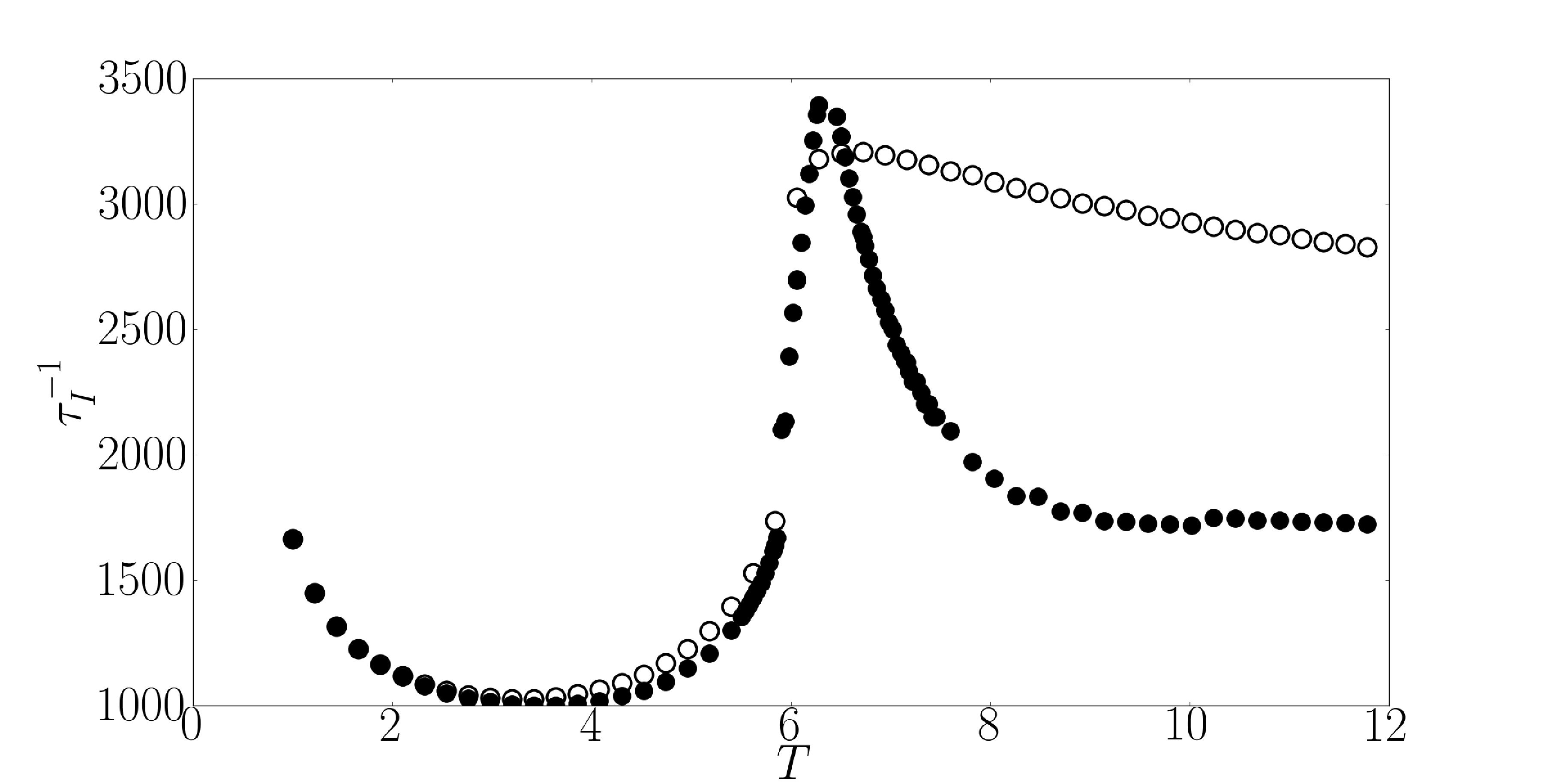,angle=0,width=9cm}}
\vspace*{8pt}
\caption{ BCC ferromagnetic film:  Inverse of relaxation time of itinerant spins $\tau_I^{-1}$ calculated with temperature-independent (white circles) and temperature-dependent  (black circles) of the lattice spins versus temperature
$T$, in zero magnetic field, with electric field $\epsilon=1$, $I_0=2$, $K_0=0.5$.}\label{RXT-F}
\end{figure}

\begin{figure}[th]
\centerline{\psfig{file=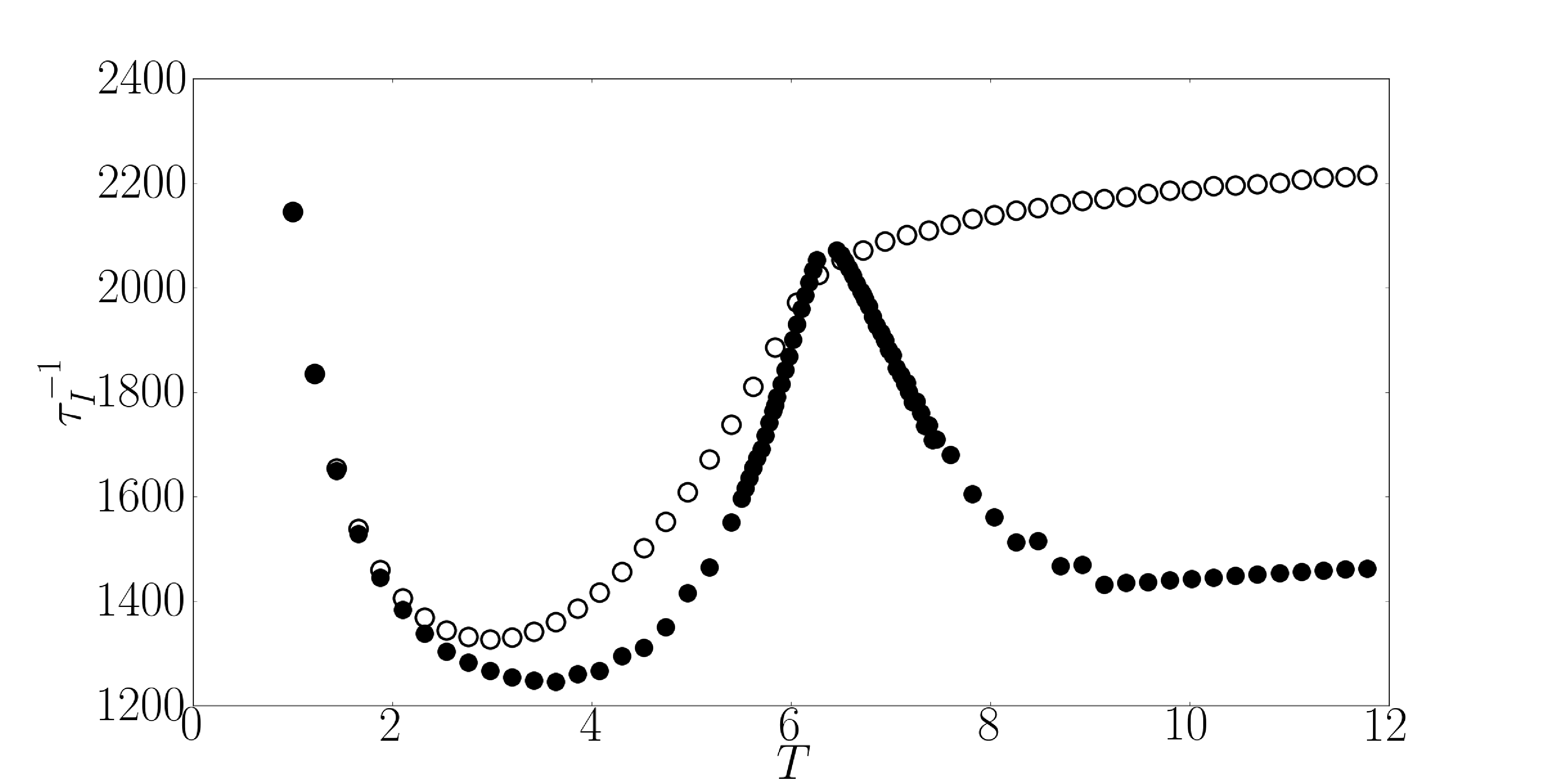,angle=0,width=9cm}}
\vspace*{8pt}
\caption{ BCC antiferromagnetic film:  Inverse of relaxation time of itinerant spins $\tau_I^{-1}$ calculated with temperature-independent (white circles) and temperature-dependent  (black circles) relaxation time of the lattice spins versus temperature
$T$, in zero magnetic field, with electric field $\epsilon=1$, $I_0=2$, $K_0=0.5$.}\label{RXT-AF}
\end{figure}

It is interesting to calculate the relaxation time $\tau_I$ of the itinerant spins.  We define $\tau_I$ in the simulations as the MC time (in unit of one MC step/spin) between two "MC collisions", namely the lapse of time between two  "rejections"  of a spin to advance. Of course this quantity is averaged over all itinerant spins and over the simulation time. Figure \ref{RXT-F} shows $\tau_I^{-1}$ obtained by simulation using $\tau_L$.  As for $R$ seen above, the temperature dependence and independence are markedly different only for $T> T_C$.  The same thing is observed for the case of antiferromagnet shown in Fig.  \ref{RXT-AF}

\begin{figure}[th]
\centerline{\psfig{file=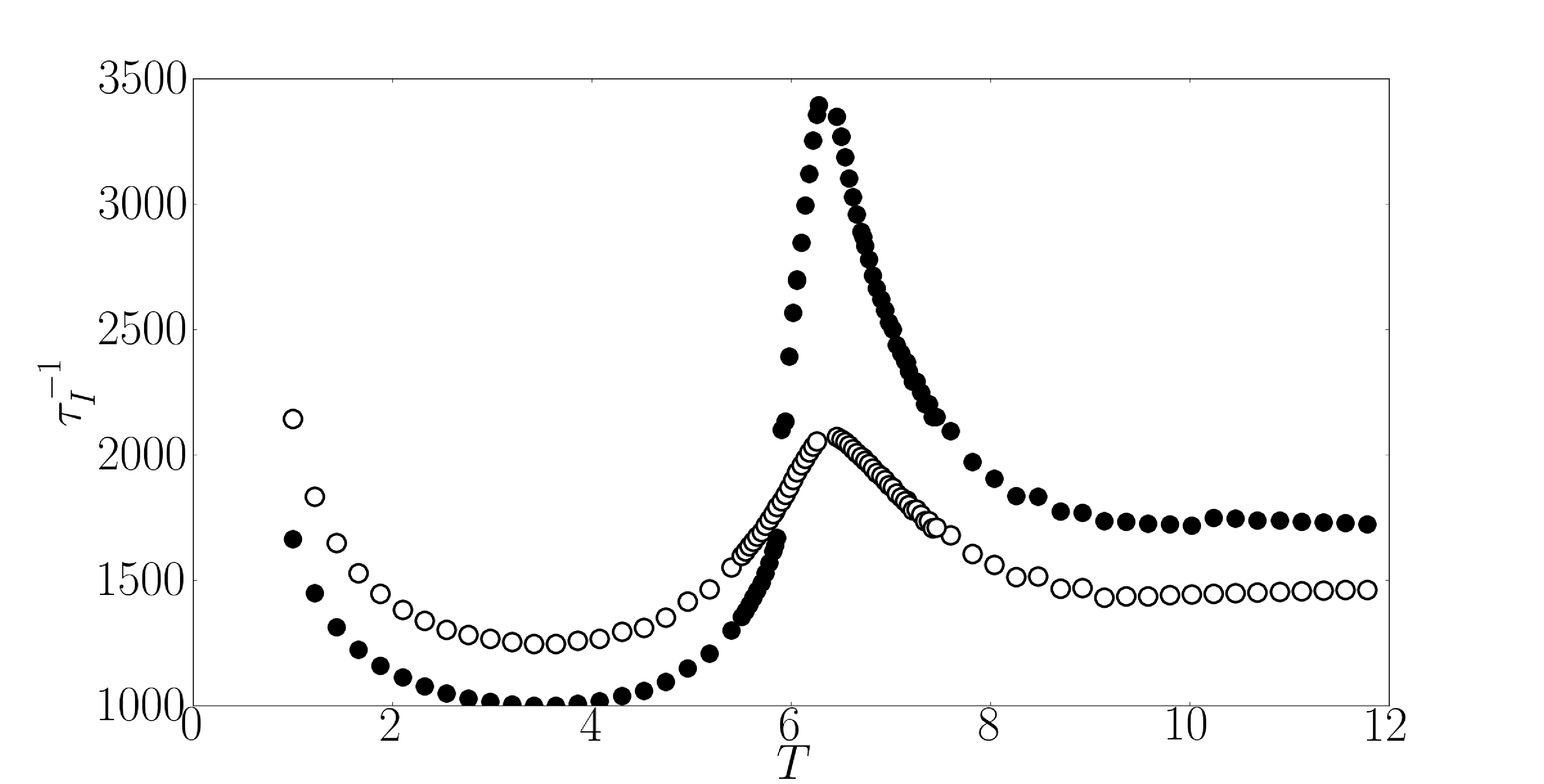,angle=0,width=9cm}}
\vspace*{8pt}
\caption{ BCC ferromagnet and antiferromagnet cases:  Inverse of relaxation time of itinerant spins $\tau_I^{-1}$ calculated with temperature-dependent relaxation time of the lattice spins for ferromagnet  (black circles) and antiferromagnet (white circles)  versus temperature $T$, in zero magnetic field, with electric field $\epsilon=1$, $I_0=2$, $K_0=0.5$.}\label{RXT-F-AF}
\end{figure}

Figure \ref{RXT-F-AF} shows $\tau_I^{-1}$ for both ferro- and antiferromagnetic cases, for comparison.  The antiferromagnet has $\tau_I^{-1}$ larger at $T<T_C$.  Note that the resistivity is proportional to $\tau_I^{-1}$.

Let us discuss the reason why the temperature dependence of the lattice relaxation time affects so strongly the shape of the spin resistivity for $T\geq T_C$.  First, we emphasize on two "empirical" rules that we observed and verified in a number of cases:
\begin{itemize}
\item a) itinerant spins move easily when they are energetically unstable. The electric field then drives them easily forward.  On the other hand, when they are "at ease" with surrounding spins, namely their energy is low, they will not move easily.  We have checked this rule by calculating their velocity as a function of their energy\cite{Magnin2}
\item b) in the case where the energy of an itinerant is low,  the move of itinerant spins depends on the energy difference between its initial and final positions.  Consider the ordered phase of the lattice:  the energy at any point is very low and the energy difference between any two points is close to zero (ordered state). So, by the MC updating criterion, the electric field dominates again the move of itinerant spins. This explains the very small resistivity at low $T$ with respect to that at high $T$ (except when $T \rightarrow 0$ where other mechanisms come to play).
\end{itemize}
For the effect of $\tau_L$, several important points are in order:

i) For $T<  T_C$ the lattice is ordered, therefore itinerant spins do not see the difference when the lattice changes its microstates more frequently or less frequently.  This explains the same values obtained for $R$ with and without temperature dependence of $\tau_L$ in ferromagnets. In antiferromagnets, one observes a small difference due to the presence of lattice down spins which act differently on up-polarized itinerant spins.

ii) For $T > T_C$, the lattice is disordered: the lattice spins are frequently flipped. Itinerant spins have to move constantly to accommodate themselves to the fluctuating environment.  Thus,  $\tau_I$ is long by definition because there are very few rejections to move.  Consequently, $R$ is small in the paramagnetic phase.

iii) Finally, it is striking to observe a strong correlation between $\tau_L$ and $\tau_I$: Since $\tau_L$ is very large in the transition region where the lattice is in the regime of critical slowing down,  itinerant spins have time to find themselves in energetically favorable positions. Once they are there they refuse to move (first rule mentioned above). As a consequence $\tau_I$ is very small (for example $\tau_I=1$ if they refuse to move at every update trial).   $R$ is thus very high. We have showed the inverse of $\tau_I$ in Figs. \ref{RXT-F}, \ref{RXT-AF} and \ref{RXT-F-AF} because $R$ is inversely proportional to $\tau_I$.   The correlation between $\tau_L$ and $\tau_I$ is thus "high $\tau_L$ corresponds to low $\tau_I$" and vice-versa.

\begin{figure}[th]
\centerline{\psfig{file=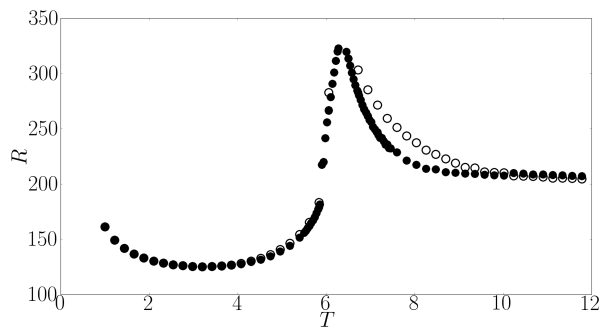,angle=0,width=9cm}}
\vspace*{8pt}
\caption{ BCC ferromagnet. Effect of the choice of the constant $A$:  i) $A=1$  (black circles), ii) $A=2$  (white circles)  versus temperature $T$, in zero magnetic field, with electric field $\epsilon=1$, $I_0=2$, $K_0=0.5$.  See text for comments.}\label{R2TAUL}
\end{figure}

Now, let us show the effect of the choice of $A$ of Eq. (\ref{tau}) in Fig. \ref{R2TAUL}. The higher $A$ (i. e. higher $\tau_L$) induces an increase of $R$ near $T_C$ but gives the same value as $T$ is far away from the critical point.   Thus, the width of $R$ at the transition temperature can serve as a measure of the relaxation time of the lattice spins.

%\begin{figure}[th]
%\centerline{\psfig{file=R_FCC_no_relax_time-var.ps,angle=0,width=9cm}}
%\vspace*{8pt}
%\caption{ FCC ferromagnet.  Temperature-independent long-time averaging  (white circles) and temperature-independent short-time averaging (black circles)  versus temperature $T$, in zero magnetic field, with electric field $\epsilon=1$, $I_0=2$, $K_0=0.5$.  See text for comments.}\label{FCC-NO-RELAX}
%\end{figure}

\section{Conclusion}\label{Concl}
In this paper, we have shown the effect of the temperature dependence of the relaxation time on the spin resistivity for both ferromagnetic and antiferromagnetic films with BCC lattice structure.

In the ferromagnetic case, the long relaxation time in the critical region compared to that of the paramagnetic phase gives rise to a sharp peak of the spin resistivity at $T_C$.   The resistivity in the low-$T$ region is insensitive to the relaxation time while in high-$T$ region, the resistivity is much smaller than that obtained with the temperature-independent relaxation time.  The same tendency is observed for the antiferromagnetic case: while the spin resistivity in the case of temperature-independent relaxation time does not show a peak at $T_C$, the extremely long relaxation in the critical region with respect to that of the paramagnetic phase gives rise to a pronounced rounded peak at $T_C$.   It is very interesting to study other systems such as spin glasses where the relaxation time is extremely long even at temperatures far below $T_C$.

\end{document}